\documentclass{article} 
\usepackage{iclr2025_conference,times}

\usepackage{amsmath,amsfonts,bm}

\def\eqref#1{equation~\ref{#1}}

\def\1{\bm{1}}

\DeclareMathAlphabet{\mathsfit}{\encodingdefault}{\sfdefault}{m}{sl}
\SetMathAlphabet{\mathsfit}{bold}{\encodingdefault}{\sfdefault}{bx}{n}

\usepackage{url}
\usepackage{array}
\usepackage{graphicx}
\usepackage{tcolorbox}
\usepackage{xcolor}
\usepackage{hyperref}
\usepackage{soul}
\usepackage{cleveref}
\usepackage{fdsymbol}
\usepackage{pifont}
\usepackage{adjustbox}

\definecolor{myblue}{rgb}{0.1, 0.1, 0.85}
\definecolor{myorange}{rgb}{0.85, 0.4, 0}
\definecolor{mygreen}{rgb}{0.05, 0.65, 0.1}
\definecolor{myred}{rgb}{0.8, 0.05, 0.05}

\newcommand*\colourcheck[1]{%
  \expandafter\newcommand\csname #1check\endcsname{\textcolor{#1}{\ding{52}}}%
}
\colourcheck{green}
\newcommand*\colourx[1]{%
  \expandafter\newcommand\csname #1x\endcsname{\textcolor{#1}{\ding{55}}}%
}
\colourx{red}
\newcommand*\yellowo{\textcolor[rgb]{1.0, 0.8, 0.1}{\textbf{O}}}

\title{Pitfalls of Evidence-Based AI Policy}

\author{Stephen Casper  \\
MIT CSAIL\\
\texttt{scasper@mit.edu}
\And
David Krueger  \\
Mila\\
\And
Dylan Hadfield-Menell  \\
MIT CSAIL
}

\iclrfinalcopy 
\begin{document}

\maketitle

\begin{quote}
    \textit{At this very moment, I say we sit tight and assess.}\\
    -- \href{https://villains.fandom.com/wiki/Janie_Orlean}{\ul{President Janie Orlean}}, \href{https://en.wikipedia.org/wiki/Don%27t_Look_Up}{\ul{Don’t Look Up}}
\end{quote}

\vspace{0.75cm}

\begin{abstract}

Nations across the world are working to govern AI. However, from a technical perspective, there is uncertainty and disagreement on the best way to do this. Meanwhile, recent debates over AI regulation have led to calls for ``evidence-based AI policy'' which emphasize holding regulatory action to a high evidentiary standard. Evidence is of irreplaceable value to policymaking. However, holding regulatory action to too high an evidentiary standard can lead to systematic neglect of certain risks. In historical policy debates (e.g., over tobacco ca. 1965 and fossil fuels ca. 1985) ``evidence-based policy'' rhetoric is also a well-precedented strategy to downplay the urgency of action, delay regulation, and protect industry interests. Here, we argue that if the goal is evidence-based AI policy, the first regulatory objective must be to actively facilitate the process of identifying, studying, and deliberating about AI risks. We discuss a set of 15 regulatory goals to facilitate this and show that Brazil, Canada, China, the EU, South Korea, the UK, and the USA all have substantial opportunities to adopt further evidence-seeking policies.

\end{abstract}


\tableofcontents


\section{How do We Regulate Emerging Tech?} \label{sec:section1}

Recently, debates over AI governance have been ongoing across the world. A common underlying theme is the challenge of regulating emerging technologies amidst uncertainty about the future. Even among people who strongly agree that it is important to regulate AI, there is sometimes disagreement about when and how. This uncertainty has led some experts and politicians to call for ``evidence-based AI policy.'' 

\subsection{“Nope, I’m against evidence-based policy.”}

See how awful that sounds? This highlights a troublesome aspect of how things are sometimes framed. Of course, evidence is indispensable. But there is a pitfall of holding policy action to too high an evidentiary standard:

\begin{center}
    \begin{tcolorbox}[colback=gray!6, colframe=white, boxrule=0.5pt, width=0.9\textwidth]
        \large
        \textbf{Postponing regulation that enables more transparency and accountability on grounds that it's ``not evidence-based'' is counterproductive.}
    \end{tcolorbox}
\end{center}

As we will argue, focusing too much on getting evidence before we act can paradoxically make it harder to gather the information we need. 

\subsection{A Broad, Emerging Coalition}

Recently, there have been a number of calls for evidence-based AI policy. For example, several California congressmembers and Governor Gavin Newsom recently argued against an AI regulatory bill in California by highlighting that it was motivated by mitigating future risks that have not been empirically observed: 

\begin{tabular}{p{6.5cm}p{0.01cm}p{6.5cm}}
    \textit{There is little scientific evidence of harm of ‘mass casualties or harmful weapons created’ from advanced models.} & & \textit{[Our] approach…must be based on empirical evidence and science…[we need] Al risk management practices that are rooted in science and fact.} \\
    -- Zoe Lofgren et al. in an \href{https://democrats-science.house.gov/imo/media/doc/2024-08-15%20to%20Gov%20Newsom_SB1047.pdf}{\ul{open letter}} to Gavin Newsom & & -- Gavin Newsom in \href{https://www.gov.ca.gov/wp-content/uploads/2024/09/SB-1047-Veto-Message.pdf}{\ul{his veto}} of bill \href{https://leginfo.legislature.ca.gov/faces/billNavClient.xhtml?bill_id=202320240SB1047}{\ul{SB 1047}}
\end{tabular}\\

Others in academia have echoed similar philosophies of governing AI amidst uncertainty. For example, in their book AI Snake Oil \citep{narayanan2024ai}, Princeton researchers Arvind Narayanan and Sayash Kapoor claim:

\begin{quote}
    \textit{The whole idea of estimating the probability of AGI risk is not meaningful…We have no past data to calibrate our predictions.}\\
    -- \citet{narayanan2024ai}, AI Snake Oil 
\end{quote}

They follow this with \href{https://www.aisnakeoil.com/p/ai-existential-risk-probabilities}{\ul{an argument}} against the precautionary principle \citep{taleb2014precautionary}, arguing that speculative estimates of future AI risks are not sufficient to warrant current action. 

Meanwhile, Jacob Helberg, a senior adviser at the Stanford University Center on Geopolitics and Technology, has argued that there just isn’t enough evidence of AI discrimination to warrant policy action.  

\begin{quote}
    \textit{This is a solution in search of a problem that really doesn't exist…There really hasn’t been massive evidence of issues in AI discrimination.}\\
    -- Jacob Helberg on \href{https://www.wired.com/story/donald-trump-ai-safety-regulation/}{\ul{priorities for the current presidential administration}} 
\end{quote}

And Martin Casado, a partner at Andreesen Horowitz, recently argued in a post that we should hold off on taking action until we know the marginal risk:

\begin{quote}
    \textit{We should only depart from the existing regulatory regime, and carve new ground, once we understand the marginal risks of AI relative to existing computer systems. Thus far, however, the discussion of marginal risks with AI is still very much based on research questions and hypotheticals.}\\
    -- \citet{casado2024base}, Base AI Policy on Evidence, Not Existential Angst 
\end{quote}

And finally, the seventeen authors of a recent article titled, \textit{A Path for Science- and Evidence-Based AI Policy}, argue that:

\begin{quote}
    \textit{AI policy should be informed by scientific understanding…if policymakers pursue highly committal policy, the…risks should meet a high evidentiary standard.}\\
    -- \citet{path_for_ai_policy}, A Path for Science‑ and Evidence‑based AI Policy 
\end{quote}

Overall, the evidence-based AI policy coalition is diverse. It includes a variety of policymakers and researchers who do not always agree with each other. 
However, \textbf{this camp is generally characterized by a desire to avoid pursuing highly committal policy absent compelling evidence.}

\subsection{A Vague Agenda?}

Calls for evidence-based policy are not always accompanied by substantive recommendations. However, here we will discuss one that does. \citet{path_for_ai_policy} end their article with a set of four milestones for researchers and policymakers to pursue:\footnote{\citet{path_for_ai_policy} also call for the establishment of a registry, evaluation, red-teaming, incident reporting, and monitoring but do not specify any particular role for regulators to play in these. They also make a nonspecific call for policymakers to broadly invest in risk analysis research and to investigate transparency requirements. }

\begin{quote}
    \textbf{Milestone 1:} A taxonomy of risk vectors to ensure important risks are well-represented\\
    \textbf{Milestone 2:} Research on the marginal risk of AI for each risk vector\\
    \textbf{Milestone 3:} A taxonomy of policy interventions to ensure attractive solutions are not missed\\
    \textbf{Milestone 4:} A blueprint that recommends candidate policy responses to different societal conditions
\end{quote}

These milestones are extremely easy to agree with. Unfortunately, they are also likely to be unworkably vague. It is unclear what it would mean for them to be accomplished. On the contrary, it is not hard to argue that existing reports reasonably meet them. For example, the AI Risk Repository \citep{slattery2024ai} predates \citet{path_for_ai_policy} and offers a meta-review, taxonomy, and living database of AI risks discussed in the literature. If this does not offer a workable taxonomy of risks (Milestone 1), it seems unclear what would.\footnote{For milestone 2, most relevant research is domain-specific; consider \citet{metta2024generativeaicybersecurity}, \citet{sandbrink2023artificial}, \citet{musser2023cost}, and \citet{cazzaniga2024gen} as examples. Note, however, that forecasting future marginal risks will always be speculative to some degree. See also \citet{bengio2025international}. Meanwhile, milestones 3 and 4 essentially describe the first and second stages of the AI regulation process, so existing regulatory efforts already are working on these (e.g., \citealp{arda2024taxonomy}).}

These milestones are an encouraging call to actively improve our understanding. However, absent more precision, we worry that similar arguments could be misused as a form of tokenism to muddy the waters and stymie policy action. 
In the rest of this paper, we will argue that holding regulatory action to too high an evidentiary standard can paradoxically make it harder to gather the information that we need for beneficial AI governance. 

\section{The Evidence is Biased}

In its pure form, science is a neutral process. But it is never done in a vacuum. Beneath the cloak of objectivity, there are subjective human beings working on problems that were not randomly selected \citep{kuhn1962structure}. There is a laundry list of biases subtly shaping the evidence produced by AI researchers. A policymaking approach that fixates on existing evidence to guide decision-making will systematically favor Western, technocratic, industry, and risk-tolerant interests

\subsection{Selective Disclosure}

In February 2023, Microsoft \href{https://blogs.microsoft.com/blog/2023/02/07/reinventing-search-with-a-new-ai-powered-microsoft-bing-and-edge-your-copilot-for-the-web/}{\ul{announced}} Bing Chat, an AI-powered web browsing assistant. It was a versatile, semi-autonomous copilot to help users browse the web. It was usually helpful, but sometimes, it went off the rails. Users found that \href{https://www.lesswrong.com/posts/jtoPawEhLNXNxvgTT/bing-chat-is-blatantly-aggressively-misaligned}{\ul{it occasionally took on shockingly angsty, deceptive, and outright aggressive personas}}. It would go so far as to \href{https://x.com/marvinvonhagen/status/1625520707768659968?lang=en}{\ul{threaten}} \href{https://x.com/sethlazar/status/1626241169754578944?s=20}{\ul{users}} chatting with it. 
It was not a tragic incident -- everyone was fine. Bing Chat was just a babbling web app that could not directly harm anyone. But it offers a cautionary tale. Right now, developers are racing to create increasingly agentic and advanced AI systems \citep{chan2023harms, casper2025ai}. If more powerful future systems go off the rails in similar ways, it could spell trouble.

Following the Bing Chat incidents, Microsoft’s public relations strategy focused on patching the issues and moving on. To the dismay of many AI researchers, Microsoft never published a public report on the incident. If Microsoft had nothing but humanity’s best interests at heart, it could substantially help researchers by reporting on the technical and institutional choices that led to Bing Chat’s behaviors. However, it’s just not in their public relations interests to do so. 

The lack of transparency around Bing Chat is no isolated occurrence. Historically, AI research and development has been a very open process. For example, code, models, and methodology behind most state-of-the-art AI systems were broadly available pre-2020. More recently, however, developers like Microsoft have been exercising more limited and selective transparency \citep{Bommasani2024TheFM}. \textbf{Due to a lack of accountability in the tech industry, some lessons remain simply out of reach. There is a mounting crisis of transparency in AI when it is needed the most.} 

\subsection{Easy- vs. Hard-to-Measure Impacts}

The scientific process may be intrinsically neutral, but not all phenomena are equally easy to study. Most of the downstream societal impacts of AI are difficult to accurately predict in a laboratory setting. The resulting knowledge gap biases purely evidence-based approaches to neglect some issues simply because they are difficult to study. 

\begin{quote}
    \textit{Thoroughly assessing downstream societal impacts requires nuanced analysis, interdisciplinarity, and inclusion…there are always differences between the settings in which researchers study AI systems and the ever-changing real-world settings in which they will be deployed.}\\
    -- \citet{bengio2024international}, International Scientific Report on the Safety of Advanced AI, Interim Report
\end{quote}

\textbf{Differences in the measurability of different phenomena can cause pernicious problems to be neglected.} For instance, compare explicit and implicit social biases in modern language models. Explicit biases from LLMs are usually easy to spot. For example, it is relatively easy to train a language model against expressing harmful statements about a demographic group. But even when we do this to language models, they still consistently express more subtle biases in the language and concept associations that they use to characterize different people \citep{wan2023kelly, wan2024white, bai2024measuring, hofmann2024dialect}.

Meanwhile, \textit{benchmarks} are the backbone of research progress in AI \citep{patterson2012technical, hendrycks2022bird}. For example, tests like GPQA \citep{rein2023gpqa} actively serve to guide progress on language model capabilities. Many of the benchmarks used in AI research are designed with the hope that they can help us understand downstream societal impacts. However, the strengths of benchmarks are also their weaknesses. Standardized, simplified, and portable measures of system performance often make for poor proxies to study real-world impacts \citep{raji2021ai}. For example, in a systematic study of benchmarks designed to assess harmlessness in AI systems, \citet{ren2024safetywashing} found that many existing benchmarks intended to evaluate these qualities were, in practice, more reflective of a model’s general capabilities than anything else.

\subsection{Precedented vs. Unprecedented Impacts}

In the history of safety engineering, many major system failures follow a certain story \citep{dekker2019foundations}. It starts off with some system – e.g., a dam, bridge, power plant, oil rig, building, etc – that functions normally for a long time. At first, this is accompanied by direct evidence of benefits and no evidence of major harms which can lull engineers into a false sense of security. But tragedy often strikes suddenly. For example, before the infamous 1986 Challenger space shuttle explosion, there were 9 successful launches \citep{gebhardt20111983}, which was a factor that led engineers to neglect safety warnings before the infamous 10th launch. Things were fine, and the empirical evidence looked good until disaster struck. AI is a very powerful technology, and if it ever has a `Chernobyl moment,' a myopic focus on empirical evidence would be the exact way to lead us there. 

\subsection{Ingroups vs. Outgroups}

The AI research community does not represent humanity well. For example, AI research is dominated by White and Asian \citep{aiindex2021diversity}, English-speaking \citep{singh2024aya} men \citep{AbdullaChahal2023}. It is also relatively culturally homogenous:

\begin{quote}
    \textit{Since AI technologies are mostly conceived and developed in just a handful of countries, they embed the cultural values and practices of these countries.}\\
    -- \citet{prabhakaran2022cultural}, Cultural Incongruities in Artificial Intelligence  
\end{quote}

As a specific example, AI ethics researchers often contrast India and the West to highlight the challenges posed by cultural homogeneity in the research community. In the West, societal discussions around fairness can, of course, be very nuanced, but they reflect Western experiences and are often characterized by a focus on race and gender politics. In India, however, the axes of social disparity are different and, in many ways, more complex. For example, India has 22 official languages, a greater degree of religious conflict, and a historical Caste system. This has led researchers to argue that the AI community is systematically poised to neglect many of the challenges in India and other non-Western parts of the world \citep{sambasivan2021re, bhatt2022re, qadri2023ai}.

\subsection{The Culture \& Values of the AI Research Community}

Perhaps the most important ingroup/outgroup contrast to consider is the one between the AI research community and the rest of humanity. It is clear that AI researchers do not demographically represent the world \citep{aiindex2021diversity, singh2024aya, AbdullaChahal2023}. Meanwhile, they tend to have much more wealth and privilege than the vast majority of the rest of the world. And they tend to be people who benefit from advances in technology instead of being historically or presently marginalized by it. This prompts a serious question:

\begin{center}
    \begin{tcolorbox}[colback=gray!6, colframe=white, boxrule=0.5pt, width=0.9\textwidth]
        \large
        \textbf{Is the AI research community prepared to put people over profit and performance?}
    \end{tcolorbox}
\end{center}

\begin{figure}[h!]
    \centering
    \includegraphics[width=0.65\linewidth]{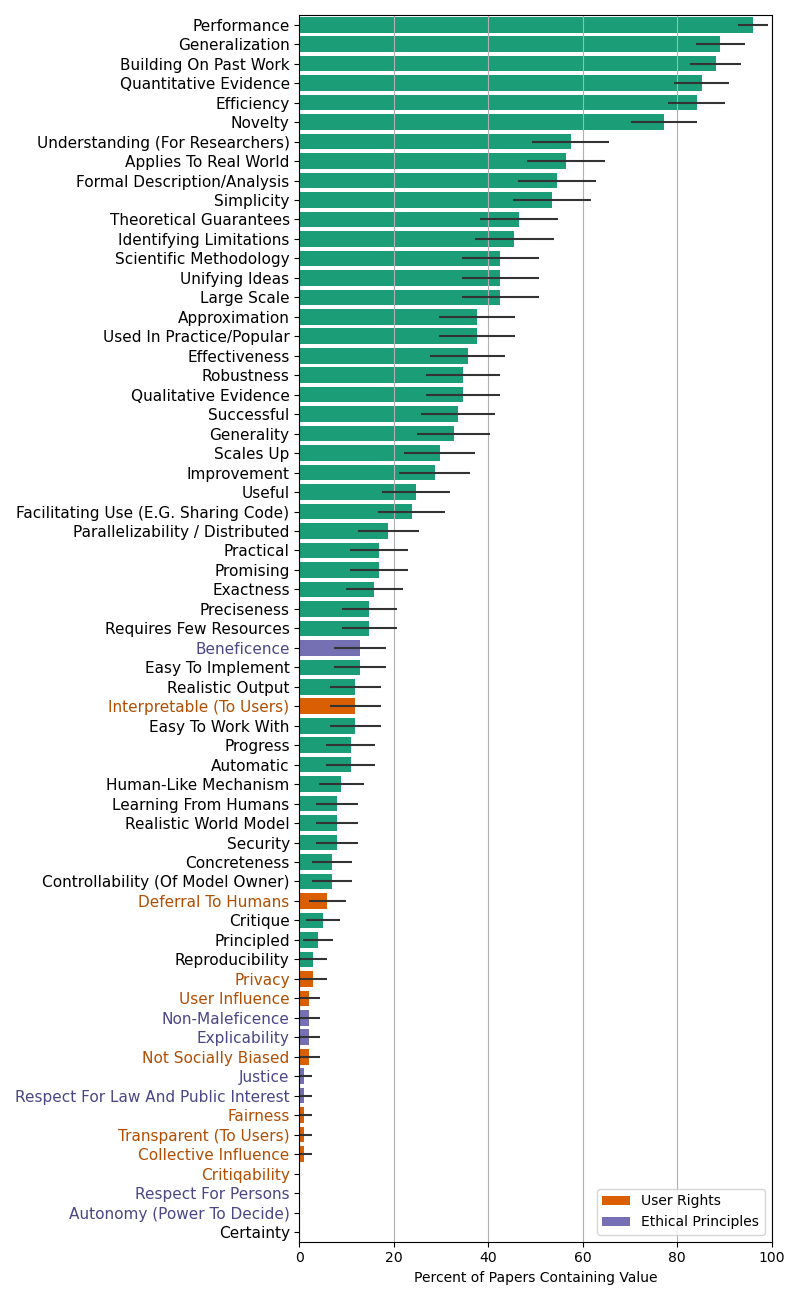}
    \caption{From \citet{birhane2022values}, \textit{The Values Encoded in Machine Learning Research}. Among the values represented in prominent AI research papers, there is an overwhelming predominance of ones pertaining to technical system performance. Is the AI research community prepared to put human interests over system performance?}
    \label{fig:birhane}
\end{figure}

In their paper, \textit{The Values Encoded in Machine Learning Research}, \citet{birhane2022values} analyzed 100 prominent, influential machine learning papers from 2008 to 2019. They annotated each based on what values were reflected in the paper text. The results, shown in \Cref{fig:birhane}, revealed a red flag. 
There was an overwhelming predominance of values pertaining to system performance (green) over the other categories of user rights and ethical principles. This suggests that the AI community may be systematically predisposed to produce evidence that will disproportionately highlight the benefits of AI compared to its harms.

\subsection{Industry’s Entanglement with Research}

\begin{figure}[h!]
    \centering
    \includegraphics[width=0.9\linewidth]{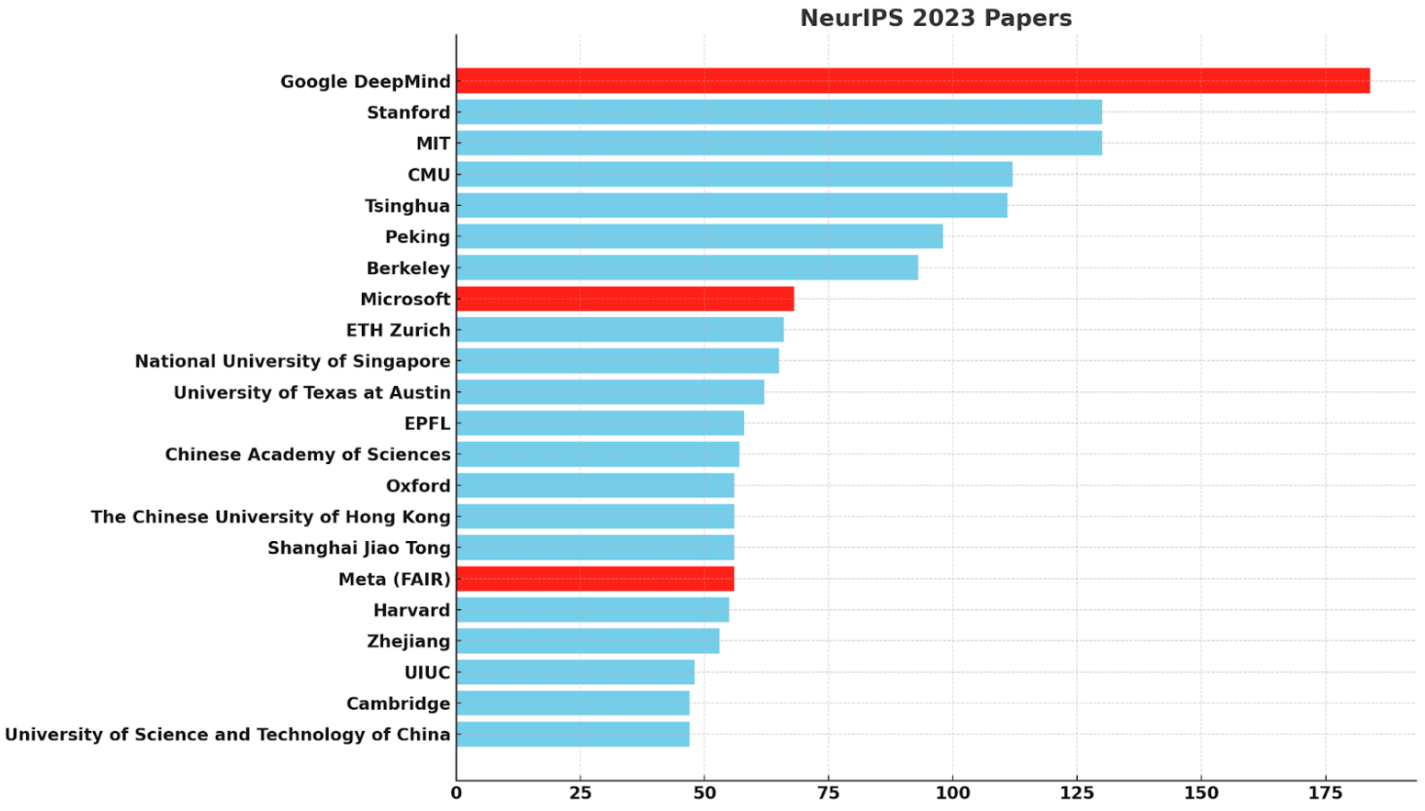}
    \caption{Paper count by organization from the \href{https://neurips.cc/Conferences/2023}{\ul{NeurIPS 2023}} conference. AI companies directly influence the evidence base.}
    \label{fig:neurips}
\end{figure}

Who is doing the AI research? Where is the money coming from? In many cases, the answer to both is the tech companies that would be directly affected by regulation. For instance, consider the \href{https://neurips.cc/Conferences/2023}{\ul{2023 NeurIPS}} conference. Google DeepMind, Microsoft, and Meta all ranked in the top 20 organizations by papers accepted (\Cref{fig:neurips}). 

\begin{figure}[h!]
    \centering
    \raisebox{0.2\height}{\includegraphics[width=0.48\linewidth]{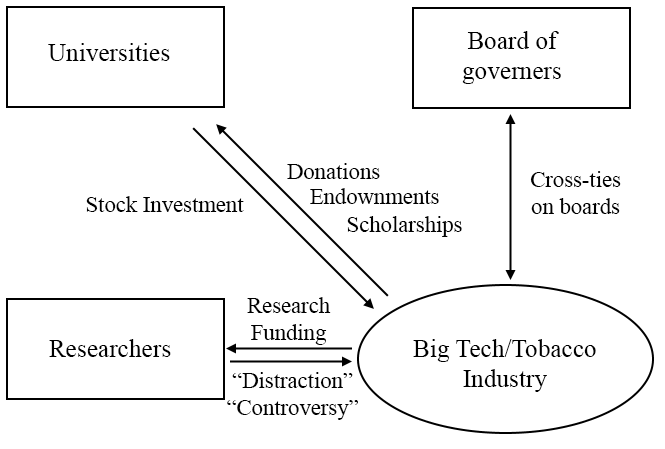}}    \includegraphics[width=0.48\linewidth]{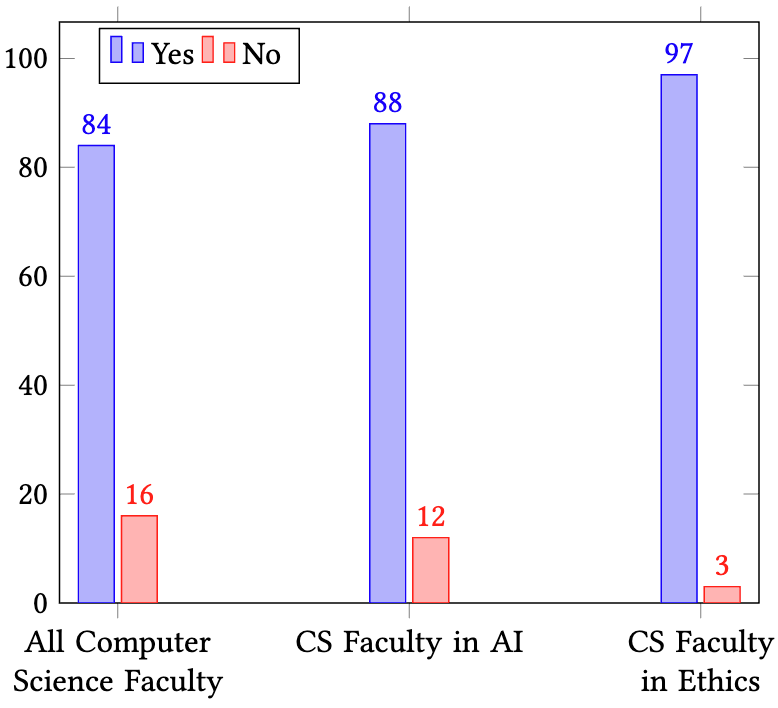}
    \caption{From \citet{Abdalla2020TheGH} (Left) There are striking similarities between the anti-regulatory influences of Big Tobacco on public health research and Big Tech on AI research. (Right) The large majority of academic CS faculty have at some point received direct funding/awards from Big Tech or have been employed by Big Tech.}
    \label{fig:abdalla}
\end{figure}

The reach of industry labs into the research space involves more than just papers. \textbf{AI academia’s entanglement with industry runs deep:}

\begin{quote}
    \textit{Imagine if, in mid-December of 2019, over 10,000 health policy researchers made the yearly pilgrimage to the largest international health policy conference in the world. Among the many topics discussed…was how to best deal with the negative effects of increased tobacco usage…Imagine if many of the speakers who graced the stage were funded by Big Tobacco. Imagine if the conference itself was largely funded by Big Tobacco.}\\
    -- A discussion alluding to the \href{https://neurips.cc/Conferences/2019}{\ul{NeurIPS 2019 conference}} from \citet{Abdalla2020TheGH}, The Grey Hoodie Project Big Tobacco, Big Tech, and the threat on academic integrity  
\end{quote}

When a powerful industry is facing regulation, it is in its interest to pollute the evidence base and public discussion around it in order to deny risks and delay action. \textbf{A key way that this manifests is with assertions that we need more evidence and consensus before we act.} 

\begin{quote}
    \textit{Is it any wonder that those who benefit the most from continuing to do nothing emphasize the controversy among scientists and the need for continued research?}\\
    -- \citet{giere2006understanding}, Understanding Scientific Reasoning 
\end{quote}

A common `\textbf{Deny and Delay Playbook}' has been used in historical policy debates to delay meaningful regulation until long after it was needed. A common story has played out in debates around tobacco, acid rain, the ozone layer, and climate change \citep{oreskes2010merchants}. In each case, industry interests pushed biased science to cast doubt on risks and made noise in the media about how there just wasn’t enough evidence to act yet. This represents a misuse of the scientific process. Of course, all scientific theories are tentative and subject to criticism – this is exactly why science is so useful. But doubtmongering can be abused against the public interest. Calls for evidence-based policy are often used more as a rhetorical pretext for kicking the can down the road than as a constructive, prescriptive proposal for action \citep{cairney2021evidence}.

\begin{quote}
    \textit{Any evidence can be denied by parties sufficiently determined, and you can never prove anything about the future; you just have to wait and see.}\\
    -- \citet{oreskes2010merchants}, Merchants of Doubt
\end{quote}

To illustrate this, we invite the reader to speculate about which of these quotes came from pundits recently discussing AI regulation and which came from merchants of doubt for the tobacco and fossil fuel industries. Who said what? 
Answers in footnote.\footnote{(Left) A cancer doctor \href{https://tinyurl.com/44k77xuw}{\ul{testifying}} to the US congress in 1965 on tobacco and public health. (Middle) Zoe Lofgren and other representatives in a 2024 \href{https://tinyurl.com/rvkp52ew}{\ul{open letter}} to Gavin Newsom on AI regulation. (Right) Fred Singer in a \href{https://pubs.acs.org/doi/pdf/10.1021/es00078a607}{\ul{paper}} arguing against climate action in 1990.}

\medskip

\begin{tabular}{p{4cm}p{0.01cm}p{4cm}p{0.01cm}p{4cm}}
    \textit{There is no need of going off [without a thorough understanding] and then having to retract…We should take no action unless it can be supported by reasonably positive evidence.} & & \textit{In addition to its misplaced emphasis on hypothetical risks, we are also concerned that [redacted] could have unintended consequences [on U.S. competitiveness]...It may be the case that the risks posed by [redacted] justify this precaution. But current evidence suggests otherwise.} & & \textit{The scientific base for [redacted] includes some facts, lots of uncertainty, and just plain ignorance; it needs more observation…There is also major disagreement…The scientific base for [redacted] is too uncertain to justify drastic action at this time.} \\
\end{tabular}\\

To see a potential example of Big Tech entangled with calls for “evidence-based AI policy,” we can scrutinize the authors behind \citet{path_for_ai_policy}: A Path for Science‑ and Evidence‑based AI Policy (discussed above in \Cref{sec:section1}). By our analysis, five of its seventeen authors have undisclosed for-profit industry affiliations.\footnote{Li, Pineau, Varoquaux, Stoica, Liang} These include the vice president of AI research at \href{https://www.meta.com/}{\ul{Meta}} and cofounders of \href{https://www.worldlabs.ai/}{\ul{World Labs}}, \href{http://together.ai}{\ul{Together.ai}}, \href{https://www.databricks.com/}{\ul{Databricks}}, \href{https://www.anyscale.com/}{\ul{Anyscale}}, and \href{https://probabl.ai/}{\ul{:probabl}}, each of which might be affected by future AI regulations.\footnote{The original version of the article did not contain any disclaimers about omitted author affiliations. However, it was updated in October 2024 to disclaim that ``Several authors have unlisted affiliations in addition to their listed university affiliation. This piece solely reflects the authors' personal views and not those of any affiliated organizations.'' However, these conflicts of interest are still not disclosed.} Not disclosing conflicts of interest in an explicitly political article does not meet generally accepted \href{https://www.acs.org/content/dam/acsorg/about/governance/committees/ethics/conflict-of-interest-10-2.pdf}{\ul{standards for ethical disclosure in research}}. We argue that in this case in particular, a policymaker reading \textit{A Path for Science‑ and Evidence‑based AI Policy} \citep{path_for_ai_policy} might interpret it very differently if it were clear that some of the authors had industry conflicts of interest. 

\section{Lacking Evidence as a Reason to Act}

So if the evidence is systematically biased, what do we do? How do we get more, better evidence? 

\subsection{Substantive vs. Process Regulation}

We argue that the need to more thoroughly understand AI risks is a reason to pass regulation--not to delay it. To see this, we first highlight the distinction between ``substantive'' regulation and ``process'' regulation. For our purposes, we define them as such:

\medskip

\begin{tabular}{p{6.5cm}p{0.01cm}p{6.5cm}}
    \textbf{Substantive} regulation limits \textbf{what} things developers can do with their AI systems. & & \textbf{Process} regulation limits \textbf{how} developers do what they do with their AI systems.\\
\end{tabular}\\

These two categories do not only apply to AI. In the food industry, for example, ingredient bans are substantive regulations while requirements for nutrition facts are process regulations. Process regulations usually pose significantly lower burdens and downsides than substantive ones. The key reason why this distinction is important is that, as we will argue:

\begin{center}
    \begin{tcolorbox}[colback=gray!6, colframe=white, boxrule=0.5pt, width=0.9\textwidth]
        \large
        \textbf{A limited scientific understanding can be a legitimate (but not necessarily decisive) argument to postpone substantive regulation. But the exact opposite applies to process regulation.}
    \end{tcolorbox}
\end{center}

Depending on whether we are considering substantive or process regulation, the argument can go different ways. To see an example, let’s consider some recent discussions on cost and compute thresholds in AI regulations.  

\subsection{ Defense of Compute \& Cost Thresholds in AI Regulation}

Some AI policy proposals set cost and compute thresholds such that, if a system’s development surpasses these, it would be subject to specific requirements. Some researchers have duly pointed out that there are hazards associated with this; cost and compute can be poor proxies for societal risk \citep{Hooker2024OnTL, heim2024training}. 

These are important and needed points about the limitations of cost and compute thresholds. For example, suppose that we are considering substantive regulations that prevent deploying certain models in certain ways. In this case, we would need careful cost-benefit analysis and the ability to adapt regulatory criteria over time. But until we have government agencies who are capable of performing high-quality evaluations of AI systems’ risks, thresholds may be the only tenable proxy available \citep{heim2024training}. 

\textbf{In the case of process regulation, there is often a lack of substantial downside.} For example, consider policies that require developers to register a system with the government if its development exceeds a cost or compute threshold. Compared to inaction, the upside is a significantly increased ability of the government to monitor the frontier model ecosystem. As for the downside? Sometimes, certain companies will accidentally be required to do more paperwork and be more transparent than regulators may have intended. Compared to the laundry list of societal-scale risks from AI \citep{slattery2024ai}, we argue that this type of risk is practically negligible. 

\section{We Can Pass Evidence-Seeking Policies Now}

It is important to understand the role of process regulation in helping us to get evidence, especially since governments often tend to underinvest in evidence-seeking during institutional design \citep{Stephenson2011InformationAA}. In contrast to vague calls for more research, we argue that a truly evidence-based approach to AI policy is one that proactively helps to produce more information. 

\begin{center}
    \begin{tcolorbox}[colback=gray!6, colframe=white, boxrule=0.5pt, width=0.9\textwidth]
        \large
        \textbf{If we want ``evidence-based'' AI policy, our first regulatory goal must be producing evidence. We don’t need to wait before passing process-based, risk-agnostic AI regulations to get more actionable information.}
    \end{tcolorbox}
\end{center}

\subsection{15 Evidence-Seeking AI Policy Objectives}

Here, we outline a set of AI regulatory goals related to \textcolor{myblue}{\textbf{institutions}}, \textcolor{myorange}{\textbf{documentation}}, \textcolor{mygreen}{\textbf{accountability}}, and \textcolor{myred}{\textbf{risk-mitigation practices}} designed to produce more evidence. Each is process-based and fully risk-agnostic. We argue that the current lack of evidence about AI risks is not a reason to delay these, but rather, a key reason why they are useful.

\begin{enumerate}
    \item \textcolor{myblue}{\textbf{AI governance institutes}}: National governments (or international coalitions) can create AI governance institutes to research risks, evaluate systems, and curate best risk management practices that developers are encouraged to adhere to. 
    \item \textcolor{myorange}{\textbf{Model registration}}: Developers can be required to register \citep{McKernon2024AIMR} frontier systems with governing bodies (regardless of whether they will be externally deployed).
    \item \textcolor{myorange}{\textbf{Model specification and basic info}}: Developers can be required to document intended use cases and behaviors (e.g., \citealp{openai_model_spec}) and basic information about frontier systems such as scale. 
    \item \textcolor{myorange}{\textbf{Internal risk assessments}}: Developers can be required to conduct and report on internal risk assessments of frontier systems. 
    \item \textcolor{myorange}{\textbf{Independent third-party risk assessments}}: Developers can be required to have an independent third-party conduct and produce a report (including access, methods, and findings) on risk assessments of frontier systems \citep{Raji2022OutsiderOD, anderljung2023towards, Casper2024BlackBoxAI}. They can also be required to document if and what “safe harbor” policies they have to facilitate independent evaluation and red-teaming \citep{longpre2024safe}.
    \item \textcolor{myorange}{\textbf{Plans to minimize risks to society}}: Developers can be required to produce a report on risks \citep{slattery2024ai} posed by their frontier systems and risk mitigation practices that they are taking to reduce them. 
    \item \textcolor{myorange}{\textbf{Post-deployment monitoring reports}}: Developers can be required to establish procedures for monitoring and periodically reporting on the uses and impacts of their frontier systems.
    \item \textcolor{myorange}{\textbf{Security measures}}: Given the challenges of securing model weights and the hazards of leaks \citep{nevo2024securing}, frontier developers can be required to document high-level non-compromising information about their security measures (e.g., \citealp{anthropic2024rsp}).
    \item \textcolor{myorange}{\textbf{Compute usage}}: Given that computing power is key to frontier AI development \citep{sastry2024computing}, frontier developers can be required to document their compute resources including details such as the usage, providers, and the location of compute clusters.  
    \item \textcolor{myorange}{\textbf{Shutdown procedures}}: Developers can be required to document if and which protocols exist to shut down frontier systems that are under their control. 
    \item \textcolor{mygreen}{\textbf{Documentation availability}}: All of the above documentation can be made available to the public (redacted) and AI governing authorities (unredacted). 
    \item \textcolor{mygreen}{\textbf{Documentation comparison in court}}: To incentivize a race to the top where frontier developers pursue established best safety practices, courts can be given the power to compare documentation for defendants with that of peer developers.
    \item \textcolor{myred}{\textbf{Labeling AI-generated content}}: To aid in digital forensics, content produced from AI systems can be labeled with metadata, watermarks, and notices. 
    \item \textcolor{myred}{\textbf{Whistleblower protections}}: Regulations can explicitly prevent retaliation and offer incentives for whistleblowers who report violations of those regulations.
    \item \textcolor{myred}{\textbf{Incident reporting}}: Frontier developers can be required to document and report on substantial incidents in a timely manner. 
\end{enumerate}

\subsection{Ample Room for Progress}

As we write this in February 2025, parallel debates over AI safety governance are unfolding across the world. There are a number of notable existing and proposed policies. 

\begin{itemize}
    \item \textbf{Brazil} proposed Bill No. 2338 of 2023 \citep{Bill2338} (proposed) on regulating the use of Artificial Intelligence, including algorithm design and technical standards. It passed the Brazilian senate in December 2024 but is pending further action before becoming law.
    \item \textbf{Canada} recently established an \href{https://ised-isde.canada.ca/site/ised/en/canadian-artificial-intelligence-safety-institute}{\ul{AI Safety Institute}} (exists), and its AI and Data Act \citep{AIDAct} (failed) recently failed in House of Commons Committee.
    \item \textbf{China} has enacted its Provisions on the Administration of Deep Synthesis Internet Information Services \citep{DeepSynthesisProvisions} (enacted), Provisions on the Management of Algorithmic Recommendations in Internet Information Services \citep{AlgorithmicRecommendationsProvisions} (enacted), and Interim Measures for the Management of Generative AI Services \citep{GenerativeAIInterimMeasures} (enacted). There are also working drafts of a potential future `The Model Artificial Intelligence Law' \citep{ModelAILaw} (proposed).
    \item In the \textbf{European Union}, the EU AI Act \citep{euaiact2024} (enacted) was passed in March 2024, and a large undertaking to design specific codes of practice \href{https://digital-strategy.ec.europa.eu/en/news/kick-plenary-general-purpose-ai-code-practice-took-place-online}{\ul{is underway}}.
    \item \textbf{South Korea} passed the Act on the Development of Artificial Intelligence and Establishment of Trust (AI Basic Act) \citep{KoreaBill2025} in December 2024. 
    \item The \textbf{UK}’s \href{https://www.aisi.gov.uk/}{\ul{AI Security Institute}} (exists) is currently building capacity and partnerships to evaluate risks and establish best risk-management practices. Thus far, the UK’s approach to AI regulation has been non-statutory (but new draft legislation may exist within a few months).
    \item In the \textbf{United States}, Donald Trump overturned Executive Order 14110 \citep{ExecutiveOrder14110} after assuming office in January 2025. This may or may not lead the \href{https://www.nist.gov/aisi}{\ul{US AI Safety Institute}} (exists) to be shut down. It also might permanently stall a potential policy \citep{federal_register_ai_reporting_2024} on model and compute reporting that the Department of Commerce proposed in response to the executive order. Meanwhile, the AI Advancement and Reliability Act \citep{HR9497} (proposed) was drafted last congress and \href{https://insideaipolicy.com/ai-daily-news/rep-obernolte-plans-reintroduce-bill-codifying-ai-safety-institute}{\ul{will be re-introduced}} this congress. Finally, the Future of AI Innovation Act \citep{S4178} (drafted) and the Preserving American Dominance in Artificial Intelligence Act \citep{S5616} (drafted) were also introduced last congress. However, as of February 2025, they are currently simply drafts. 
\end{itemize}

So how are each of these countries faring?

\begin{table}[h!]
\begin{adjustbox}{center}
    \centering
    \renewcommand{\arraystretch}{1.3}
    \begin{tabular}{lccccccc}
        & \textbf{Brazil} & \textbf{Canada} & \textbf{China} & \textbf{EU} & \textbf{Korea} & \textbf{UK} & \textbf{USA} \\
        \textbf{\textcolor{myblue}{1. AI governance institutes}} & \yellowo * & \greencheck & \yellowo * & \greencheck & \greencheck & \greencheck & \greencheck \\
        \textbf{\textcolor{myorange}{2. Model registration}} & \redx & \redx & \greencheck & \greencheck & \greencheck & \redx & \yellowo * \\
        \textbf{\textcolor{myorange}{3. Model specification and basic info}} & \yellowo * & \redx & \yellowo & \greencheck & \yellowo & \redx & \redx \\
        \textbf{\textcolor{myorange}{4. Internal risk assessments}} & \greencheck * & \redx & \yellowo & \greencheck & \greencheck & \redx & \redx \\
        \textbf{\textcolor{myorange}{5. Independent 3rd-party risk assessments}} & \redx & \redx & \yellowo & \yellowo & \yellowo & \redx & \redx \\
        \textbf{\textcolor{myorange}{6. Plans to minimize risks to society}} & \yellowo * & \redx & \yellowo & \greencheck & \yellowo & \redx & \redx \\
        \textbf{\textcolor{myorange}{7. Post-deployment monitoring reports}} & \redx & \redx & \redx & \greencheck & \redx & \redx & \redx \\
        \textbf{\textcolor{myorange}{8. Security measures}} & \redx & \redx & \yellowo & \greencheck & \redx & \redx & \redx \\
        \textbf{\textcolor{myorange}{9. Compute usage}} & \redx & \redx & \redx & \yellowo & \redx & \redx & \yellowo * \\
        \textbf{\textcolor{myorange}{10. Shutdown procedures}} & \greencheck * & \redx & \redx & \yellowo & \redx & \redx & \redx \\
        \textbf{\textcolor{mygreen}{11. Documentation availability}} & \redx & \redx & \yellowo & \yellowo & \yellowo & \redx & \redx \\
        \textbf{\textcolor{mygreen}{12. Documentation comparison in court}} & \redx & \redx & \redx & \redx & \redx & \redx & \redx \\
        \textbf{\textcolor{myred}{13. Labeling AI-generated content}} & \redx & \redx & \greencheck & \yellowo & \yellowo & \redx & \redx \\
        \textbf{\textcolor{myred}{14. Whistleblower protections}} & \redx & \redx & \greencheck & \greencheck & \redx & \redx & \redx \\
        \textbf{\textcolor{myred}{15. Incident reporting}} & \greencheck * & \redx & \redx & \greencheck & \redx & \redx & \redx \\
    \end{tabular}
\end{adjustbox}
    \caption{\greencheck = Yes, \yellowo = Partial, \redx = No, $*$ = proposed but not enacted. There is significant room for progress across the world on passing evidence-seeking AI policy measures. See details on each row above. Note that this table represents a snapshot in time (February 2025). In the USA, we omit the two bills \citep{S4178, S5616} discussed above that were proposed last congress but have not been scheduled for re-introduction this congress.}
    \label{tab:ai_governance}
\end{table}

\subsection{The Duty to Due Diligence from Discoverable Documentation of Dangerous Deeds}

The objectives outlined above hinge on documentation. 2-10 are simply requirements for documentation, and 11-12 are accountability mechanisms to ensure that the documentation is not perfunctory. This is no coincidence. When it is connected to external scrutiny, documentation can be a powerful incentive-shaping force \citep{tomei2025ai}. Under a robust regime implementing the above the public and courts could assess the quality of developers' risk management practices. As such, this type of regulatory regime could incentivize a race to the top on risk-mitigation standards \citep{Hadfield2023RegulatoryMT}.

We refer to this phenomenon as the \textbf{Duty to Due Diligence from Discoverable Documentation of Dangerous Deeds – or the 7D effec}t. Regulatory regimes that induce this effect are very helpful for improving accountability and reducing risks. Unfortunately, absent requirements for documentation and scrutiny thereof, developers in safety-critical fields have a perverse incentive to intentionally suppress documentation of dangers. For example, common legal advice warns companies against documenting dangers in written media:

\begin{quote}
    \textit{These documents may not seem bad, but in the hands of an opposing attorney these cold hard facts can [be] used to swing a judge or a jury to their side. Often the drafters of these documents tend to believe that they are providing the company with some value to the business. For example, an engineer notices a potential liability in a design so he informs his supervisor through an email. However, the engineer’s lack of legal knowledge…may later implicate the company…when a lawsuit arises.}\\
    -- FindLaw Attorney Writers (2016), \href{https://corporate.findlaw.com/litigation-disputes/safe-communication-guidelines-for-creating-corporate-documents.html}{\ul{Safe Communication: Guidelines for Creating Corporate Documents That Minimize Litigation Risks}}
\end{quote}

We personally enjoyed the use of ``when'' and not ``if'' in this excerpt. 

Meanwhile, there is legal precedent for companies to sometimes lose court cases when they internally communicate risks through legally discoverable media such as in Grimshaw v. Ford \citep{grimshaw1981}. \textbf{So absent requirements, companies will tend to actively suppress the documentation of dangers to avoid accountability.} Meanwhile, mere voluntary transparency can be deceptive by selectively revealing information that reflects positively on the company \citep{Ananny2018SeeingWK}. Thus, we argue that a regime that requires developers to produce scrutable documentation will be key to facilitating the production of more meaningful evidence. 

\subsection{Considering Counterarguments}

\textbf{``These 15 objectives would be too burdensome for developers.''} It is true that protocols and documentation can impose burdens. However, these burdens are generally far lighter than those imposed by substantive regulations, and compliance with many of these requirements may be trivial for developers already planning to take similar actions internally. For instance, even the most potentially burdensome measures – such as risk assessments and staged deployments – are practices that major developers like \href{https://cdn.openai.com/openai-preparedness-framework-beta.pdf}{\ul{OpenAI}}, \href{https://assets.anthropic.com/m/24a47b00f10301cd/original/Anthropic-Responsible-Scaling-Policy-2024-10-15.pdf}{\ul{Anthropic}}, and \href{https://storage.googleapis.com/deepmind-media/DeepMind.com/Blog/introducing-the-frontier-safety-framework/fsf-technical-report.pdf}{\ul{Google DeepMind}} have already publicly committed to implementing.

\textbf{``It’s a slippery slope toward overreach.''} A second concern is that these 15 regulatory objectives might generate information that could pave the way for future substantive regulations or liability for developers. Regulatory and liability creep are real phenomena that can harm industry interests. However, it is important to emphasize that any progression from these objectives to future regulations or liability will ultimately depend on human decision-makers acting on evidence. Evidence is essential for society to engage in meaningful deliberation and exercise informed agency – this is the entire point of evidence-based policy. Therefore, if process-based AI regulations eventually lead to substantive regulations, it will not be because the process regulations laid an inevitable framework. It would be because the information produced by those regulations persuaded policymakers to take further action. Ultimately, we believe it would be very precarious to argue that a democratic society should be protected from its own access to information about what it is consuming. 

\section{Building a Healthier Ecosystem}

Governing emerging technologies like AI is hard \citep{Bengio2023ManagingEA}. We don’t know what is coming next. We echo the concerns of other researchers that there are critical uncertainties with the near and long-term future of AI. Anyone who says otherwise is probably trying to sell you something. So how do we go about governing AI under uncertainty? History teaches us some lessons about the importance of prescient action. 

\begin{quote}
    \textit{[Early] studies of global warming and the ozone hole involved predicting damage before it was detected. It was the prediction that motivated people to check for damage; research was intended in part to test their prediction, and in part to stimulate action before it was too late to stop...It was too soon to tell whether or not widespread and serious…damage was occurring, but the potential effects were troubling…A scientist would be in a bit of a bind: wanting to prevent damage, but not being able to prove that damage was coming…There are always more questions to be asked.}\\
    -- \citet{oreskes2010merchants}, Merchants of Doubt
\end{quote}

We often hear discussions about how policymakers need help from AI researchers to design technically sound policies. This is essential. But there is a two-way street. Policymakers can do a great deal to help researchers, governments, and society at large to better understand and react to AI risks \citep{kolt2024responsible}. 

\textbf{Process regulations can lay the foundation for more informed debates and decision-making in the future.} Right now, the principal objective of AI governance work is not necessarily to get all of the right substantive regulations in place. Instead, we argue that it is to shape the AI ecosystem to help us better identify, study, and deliberate about risks. This requires being critical of the biases shaping the evidence we see and proactively working to seek more information. Kicking the can down the road for a lack of `enough' evidence could impair policymakers' ability to take needed action. 

This lesson is sometimes painfully obvious in retrospect. For example, in the 1960s and 70s, a scientist named S.J. Green was head of research at the British American Tobacco (BAT) company. He helped to orchestrate BAT’s campaign to deny urgency and delay action on public health risks from tobacco. However, he later split with the company, and after reflecting on the intellectual and moral irresponsibility of these efforts, he remarked:

\begin{quote}
    \textit{Scientific proof, of course, is not, should not, and never has been the proper basis for legal and political action on social issues. A demand for scientific proof is always a formula for inaction and delay and usually the first reaction of the guilty. The proper basis for such decisions is, of course, quite simply that which is reasonable in the circumstance.}\\
    -- S. J. Green, \href{https://www.industrydocuments.ucsf.edu/tobacco/docs/#id=hxgc0040}{\ul{Smoking, Related Disease, and Causality}}
\end{quote}

\bigskip

\section*{Acknowledgments}

We are thankful for our discussions with Akash Wasil, Ariba Khan, Aruna Sankaranarayanan, Dawn Song, Kwan Yee Ng, Landon Klein, Rishi Bommasani, Shayne Longpre, and Thomas Woodside.

\newpage
\bibliography{bibliography}
\bibliographystyle{iclr2025_conference}

\end{document}